\documentclass[%
 reprint,
 amsmath,amssymb,
 aps,
]{revtex4-2}

\usepackage{graphicx}
\usepackage{dcolumn}
\usepackage{bm}

\usepackage{hyperref}

\usepackage{color}

\newcommand\invisible[1]{\textcolor{white}{#1}}

\newcommand\sub[1]{\mbox{\scriptsize{#1}}}

\begin{document}

\preprint{APS/123-QED}

\title{Enabling infinite $Q$ factors in absorbing optical systems}

\author{Radoslaw Kolkowski}
 \email{radoslaw.kolkowski@aalto.fi}
\author{Andriy Shevchenko}%
 \email{andriy.shevchenko@aalto.fi}
\affiliation{%
Department of Applied Physics, Aalto University, P.O.Box 13500, Aalto FI-00076, Finland
}%

\date{\today}

\begin{abstract}
Resonant optical systems have widespread applications in science and technology. However, their quality ($Q$) factors can be significantly deteriorated, if some of their parts exhibit optical absorption. Here, we show that by coupling a lossy mode of such a structure to two independent lossless modes, one can create a nonradiating and absorption-free bound state in the continuum (BIC). The $Q$ factor of such a BIC is theoretically unlimited despite interaction with an absorbing structure. We use this mechanism to design a plasmonic metasurface with $Q$ factors that are close to $10^7$ in the visible spectral range. The proposed mechanism is general and can be used to engineer ultrahigh-$Q$ resonances in various absorbing structures.
\end{abstract}

\maketitle

\section{Introduction}

Resonant optical systems are an essential part of modern science and technology, enabling the operation of lasers and providing the means for controlling various light-matter interactions. The main parameter describing such systems is the $Q$ factor, which quantifies the average amount of time each photon spends trapped inside the resonator before it either escapes or gets absorbed. Recently, the concept of bound states in the continuum (BICs) has provided an efficient strategy for eliminating the radiation loss \cite{hsu16,koshelev19__}, making it possible to realize high-$Q$ resonances in a diverse variety of non-absorbing optical systems \cite{marinica08,koshelev19}, including photonic crystals \cite{hsu13,doeleman18}, metasurfaces \cite{koshelev18,overvig20,overvig22,li22}, and individual dielectric nanoresonators \cite{rybin17,bogdanov19}. The highest $Q$ factors of BICs achieved in such non-absorbing systems are on the order of $10^5$ - $10^6$ \cite{jin19,chen22}, which is not far from the state-of-the-art photonic crystal cavities that can achieve $Q$ factors on the order of $10^7$ \cite{sekoguchi14,asano17}. Therefore, optical BICs have been considered promising for a range of applications \cite{azzam21,kupriianov19,shi22,malek22}, e.g., in laser technology \cite{ha18,kodigala17,hwang21,mohamed22,heilmann22,salerno22}  and nonlinear optics \cite{carletti18,koshelev19_,minkov19,anthur20}. 

However, structures made of absorbing materials have always been regarded as detrimental to the $Q$ factors and not suitable for ultrahigh-$Q$ systems \cite{joseph21}, unless absorption can be compensated for by gain \cite{krasnok18}. This is because  destructive interference in the far-field reduces the radiation loss, but the absorption loss remains unaffected, if not increased \cite{wang20,xiao21,saadabad21,zong23}. On the other hand, absorption can be suppressed via destructive interference in the near-field \cite{zhang12,fang15,pirruccio16}, which typically occurs under different conditions than those required by BICs. This has been the reason for relatively modest $Q$ factors, on the order of $10^2$ - $10^3$, achieved by BICs in plasmonic and hybrid plasmonic-dielectric structures~\cite{monticone17,azzam18,liang20,kolkowski20,xiang20,meudt20,sun21,deng22,aigner22,zhou22,cao23}. Consequently, absorbing materials, such as metals and some high-index semiconductors have been avoided when realizing BICs and other types of ultrahigh-$Q$ resonators, despite the fact that these materials can be very efficient in controlling optical fields~\cite{novotny11,koenderink15,kuznetsov16}. 

In this work, we discover a simple and general mechanism leading to a BIC in which \emph{both} the radiation and absorption losses are eliminated. Such a BIC can be realized by coupling the lossy oscillator to two initially lossless optical modes. The losses due to both the radiation and absorption are \emph{simultaneously suppressed} when the two lossless modes have equal resonance frequencies. There are no explicit symmetry requirements for the coupled modes as long as the coupling mechanism is preserved. This makes the presented BIC very general compared to other types of BICs that are subject to various symmetry constraints. Destructive interference of the coupled modes makes the underlying mechanism similar to the previously studied Friedrich-Wintgen BICs \cite{friedrich85} that have been observed in systems with avoided crossings~\cite{azzam18,lee20,tian20,meudt20,sun21,deng22}. Some of these systems exhibited a reduction of the absorption loss~\cite{tian20}. However, the possibility to completely eliminate the absorption loss together with the radiation loss has not been in the focus of previous works.

\begin{figure*}[hbt!]
\centering\includegraphics[width=18cm]{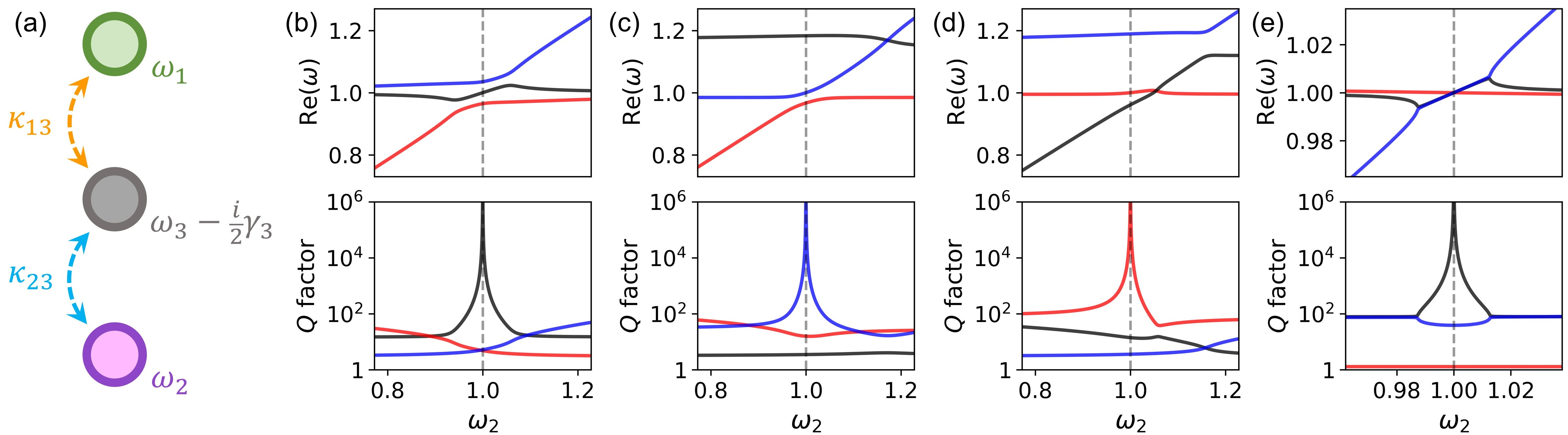}
\caption{\label{fig:1} (a) Illustration of the coupling between three oscillators described by Hamiltonian $\bm{H}$ in Eq. (\ref{eq:H}). The plots in (b)-(e) show the real parts of the complex eigenvalues $\omega$ (top) and the corresponding $Q$ factors ($Q = \frac{1}{2}\operatorname{Re}(\omega)/\operatorname{Im}(\omega)$, bottom) of the eigenstates of Hamiltonian $\bm{H}$, as functions of the resonance frequency $\omega_2$. The plots in (b) show the resonant case ($\omega_3 = \omega_1$) in which the BIC coincides with the lossy resonance, whereas (c) and (d) correspond to the off-resonant scenario ($\omega_3 = 1.15\omega_1$). The plots in (e) correspond to a strongly non-Hermitian system (with $\omega_3 = \omega_1$) in which the BIC is accompanied by a pair of exceptional points instead of an avoided crossing. In all plots, $\omega_1 = 1$, whereas the other parameters are: $\kappa_{13}=\kappa_{23}=0.075$ and $\gamma_3 = 0.2$ in (b) and (c), $\kappa_{13}=0.045$, $\kappa_{23}=0.105$ and $\gamma_3 = 0.2$ in (d), and $\kappa_{13}=\kappa_{23}=0.05$ and $\gamma_3 = 0.4$ in (e).}
\end{figure*}

Using numerical simulations, we demonstrate the existence of such BICs in a realistic physical system -- a periodic metasurface composed of a slab waveguide and an array of metal nanoparticles~\cite{christ03,rodriguez12}. By this example, we show that the proposed BICs can originate from modes of different types, such as TE and TM guided modes. In this case, the coupling results in a BIC with a mixed TE-TM polarization state. The hybrid resonances we introduce here can be perfectly lossless, even though the coupling of the modes is provided by lossy nanostructures. Our findings suggest that even strongly absorbing plasmonic or semiconductor materials can be used to construct ultrahigh-$Q$ resonant systems based on BICs. The unique properties of such materials, combined with those of BICs, can lead to superior photonic devices with new functionalities.

\section{Mechanism of loss cancellation}

Consider a system of coupled harmonic oscillators described by the following equations:
\begin{equation}
\label{eq:osc}
    \frac{d^2x_p}{dt^2}+\gamma_p\frac{dx_p}{dt}+\omega_p^2x_p-\sum_{q\neq p}\Omega^2_{pq}x_q=0,
\end{equation}
where $x_p$ is the instantaneous amplitude of a $p^{\sub{th}}$ oscillator, $\gamma_p$ the damping rate, $\omega_p$ the resonance frequency, and $\Omega_{pq}$ the coupling rate between the oscillators $p$ and $q$. Assuming the ansatz $x_p=x_{p,0}e^{-i\omega t}$ and using the approximation $|\omega_p-\omega|\ll\omega$, the above set of equations can be simplified into a linear eigenvalue problem \cite{rodriguez16}: $\mbox{det}(\bm{H}-\omega\bm{I})=0$. Here, $\bm{I}$ is the identity matrix, whereas the Hamiltonian $\bm{H}$ has the diagonal terms $\omega_p-\frac{i}{2}\gamma_p$ and off-diagonal terms $\kappa_{pq}=\frac{1}{2}\Omega^2_{pq}/\bar{\omega}$ with $\bar{\omega}\approx(\omega+\omega_p)/2$.

Now, consider a system composed of two oscillators with resonance frequencies $\omega_1$ and $\omega_2$ that are not directly coupled to each other ($\kappa_{12}=0$) and show no losses ($\gamma_{1}=\gamma_{2}=0$). Next, we introduce a third oscillator with resonance frequency $\omega_3$ and damping rate $\gamma_3$ coupled to the first and second oscillators via $\kappa_{13}$ and $\kappa_{23}$, respectively [see Fig. \ref{fig:1}(a)]. This system of oscillators is described by the Hamiltonian
\begin{equation}
\label{eq:H}
\bm{H}=\begin{pmatrix}\omega_1 & 0 & -\kappa_{13}\\
    0 & \omega_{2}  & -\kappa_{23}\\
    -\kappa_{13} & -\kappa_{23} & \omega_3-\frac{i}{2}\gamma_3\end{pmatrix}.
\end{equation}
In the context of optical resonances, the eigenstates of this Hamiltonian are quasi-normal modes \cite{alpeggiani17,lalanne18,krasnok19} associated with complex eigenvalues $\omega$, where $\mbox{Im}(\omega)$ describes the losses. A somewhat similar three-mode system has been considered as a generalization to the Friedrich-Wintgen BIC \cite{sun21,shubin22} and as a classical analog of the double electromagnetically induced transparency \cite{bai13}. Solving the eigenvalue problem yields the following relation:
\begin{eqnarray}
(\omega_1-\omega)(\omega_2-\omega)(\omega_3-\tfrac{i}{2}\gamma_{3}-\omega)=\nonumber\\\kappa^2_{23}(\omega_1-\omega)+\kappa^2_{13}(\omega_2-\omega).
\end{eqnarray}
For $\omega_{1}=\omega_{2}$, this equation can be written as
\begin{eqnarray}
(\tilde{\omega}-\omega)^2(\omega_3-\tfrac{i}{2}\gamma_{3}-\omega)=(\kappa^2_{13}+\kappa^2_{23})(\tilde{\omega}-\omega),
\end{eqnarray}
where $\tilde{\omega}=\omega_{1}=\omega_{2}$. The above relation clearly shows that the system under consideration hosts a lossless hybrid eigenstate at $\omega=\tilde{\omega}$ that is fully independent of the elements $\kappa_{13}$, $\kappa_{23}$, and $\omega_{3}-\frac{i}{2}\gamma_{3}$. Most importantly, this eigenstate is completely immune to losses expressed by $\gamma_{3}$, which may include also the absorption loss. This is in contrast to the Friedrich-Wintgen BIC emerging from two resonances with radiation losses only \cite{hsu16,friedrich85}.

The appearance of a lossless BIC in the above system is illustrated in Fig. \ref{fig:1}(b)-(e), where the eigenvalues of $\bm{H}$ are plotted as functions of the resonance frequency $\omega_2$. The BIC is present at $\omega_2=\omega_1$, regardless of the choice of parameters $\omega_3$, $\gamma_3$, $\kappa_{13}$, and $\kappa_{23}$. In particular, Fig. \ref{fig:1}(b) shows the resonant scenario ($\omega_1=\omega_3$), whereas in Fig. \ref{fig:1}(c), the lossy resonance of frequency $\omega_3$ is strongly detuned from $\omega_1$. The BIC would also emerge if the rest of the parameters was chosen differently, giving rise to hybrid modes of a qualitatively distinct character. For example, if $\kappa_{13}\neq\kappa_{23}$ or $\kappa_{13/23}\ll\gamma_3$, one can obtain the accidental degeneracies and exceptional points~\cite{zhen15,deng22} in addition to the BIC, which is illustrated in Figs. \ref{fig:1}(d) and \ref{fig:1}(e).

A relatively similar mechanism to eliminate absorption is that of the electromagnetically induced transparency \cite{fleischhauer05,liu09}. In our case, however, the loss cancellation occurs by interference of three excited modes instead of two driving fields. Similarly, in the symmetry-protected BICs~\cite{koshelev18,overvig18,overvig20_} and Friedrich-Wintgen BICs~\cite{azzam18,lee20,tian20,meudt20,sun21,deng22}, the modes can exhibit a reduced absorption loss due to their local interference in the near-field, that can be destructive to some degree~\cite{tian20,sun21}. However, the coupling mechanism presented here ensures that simultaneous cancellation of the radiation and absorption losses depends only on the frequency matching of the modes and takes place without any specific symmetry requirements, as long as the coupling mechanism is retained. 

The proposed mechanism can be implemented in various physical systems. One of them is a linear array of coupled waveguides. In a configuration of three parallel waveguides, the middle one can be lossy, while the other two lossless. If the two side waveguides are coupled only through the central waveguide, then the propagation of light in the array can be described by the same Hamiltonian as in Eq. (\ref{eq:H}). The main difference between the coupled waveguides and coupled oscillators is that, in the case of coupled waveguides, the eigenvalues $\omega$ should be interpreted as the propagation constants of the eigenmodes, and not as their eigenfrequencies. One eigenmode, in which absorption is eliminated, has an antisymmetric field distribution with zero amplitude in the middle waveguide and out-of-phase oscillation in the side waveguides \cite{maurya22}. This eigenmode would propagate without losses despite the fact that the two waveguides on the sides are coupled through the lossy waveguide.

A slightly more complex optical system that realizes the proposed mechanism would be a system of three coupled ring resonators, in which the central resonator is lossy. One may expect an eigenmode, in which light is trapped in the two lossless resonators, avoiding the lossy one. However, the coupling would still occur through the lossy resonator, locking the relative phase of the fields in the lossless resonators. The fields of the two coupled modes in the lossy resonator will disappear due to destructive interference. Other types of optical resonators can also be considered, e.g., Fabry-P\'{e}rot cavities \cite{geng20}, photonic crystal cavities \cite{yang09}, antenna-cavity hybrids \cite{doeleman16,cognee19,doeleman20}, optomechanical resonators \cite{verhagen12}, and elementary quantum oscillators, such as atoms and molecules \cite{torma14}. The presented BIC would also have analogs in physical systems beyond optics, e.g., in purely mechanical oscillators (starting with the classic example of coupled pendulums), acoustic resonators \cite{xiao17}, and electric circuits \cite{amrani22}.

\begin{figure}[hbt!]
\centering\includegraphics[width=7.5cm]{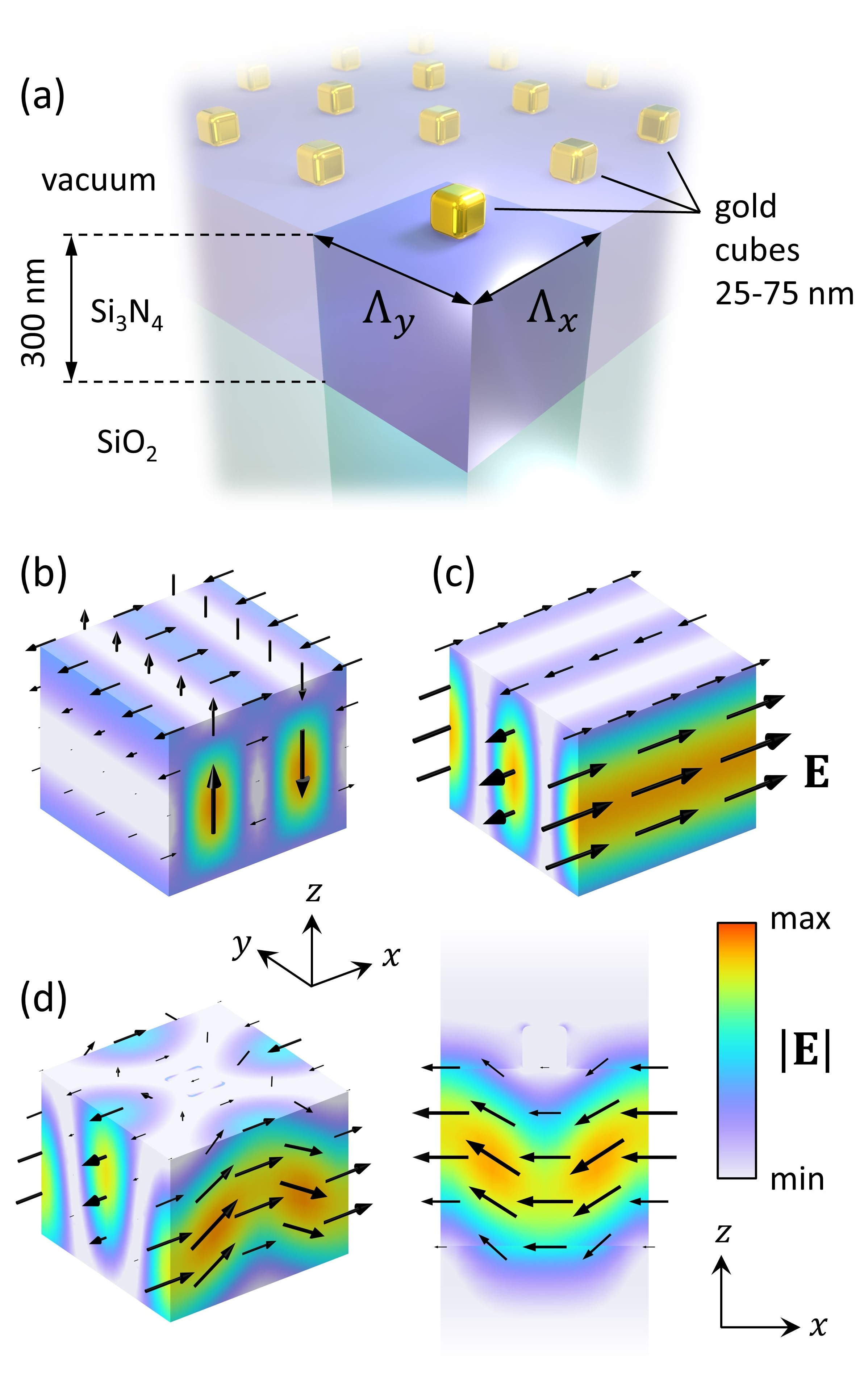}
\caption{\label{fig:2} (a) Schematic illustration of a metasurface supporting hybrid BICs. The unit cell and the two lattice periods $\Lambda_x$ and $\Lambda_y$ are highlighted in the foreground. In (b), (c), and (d), the spatial distributions of the electric field norm (color) and instantaneous electric field vector (black arrows) are shown. They were obtained using COMSOL Multiphysics. In (b) and (c), two eigenmodes of a bare Si$_3$N$_4$ waveguide at the $\Gamma$ point are presented: (b) TM mode, forming a standing wave along $x$ with a node in the center of the unit cell; (c) TE mode, forming a standing wave along $y$ with an antinode in the center of the unit cell. (d) A hybrid quasi-BIC excited in a metasurface by normally incident plane wave polarized along $x$ ($\lambda\approx$ 645.43 nm, $\Lambda_x = $ 352.5 nm, $\Lambda_y = $ 340 nm, nanocube size 75 nm): 3D view of the unit cell (on the left) and cross-cut in the $xz$-plane (on the right).}
\end{figure}

\begin{figure}[hbt!]
\centering\includegraphics[width=7.5cm]{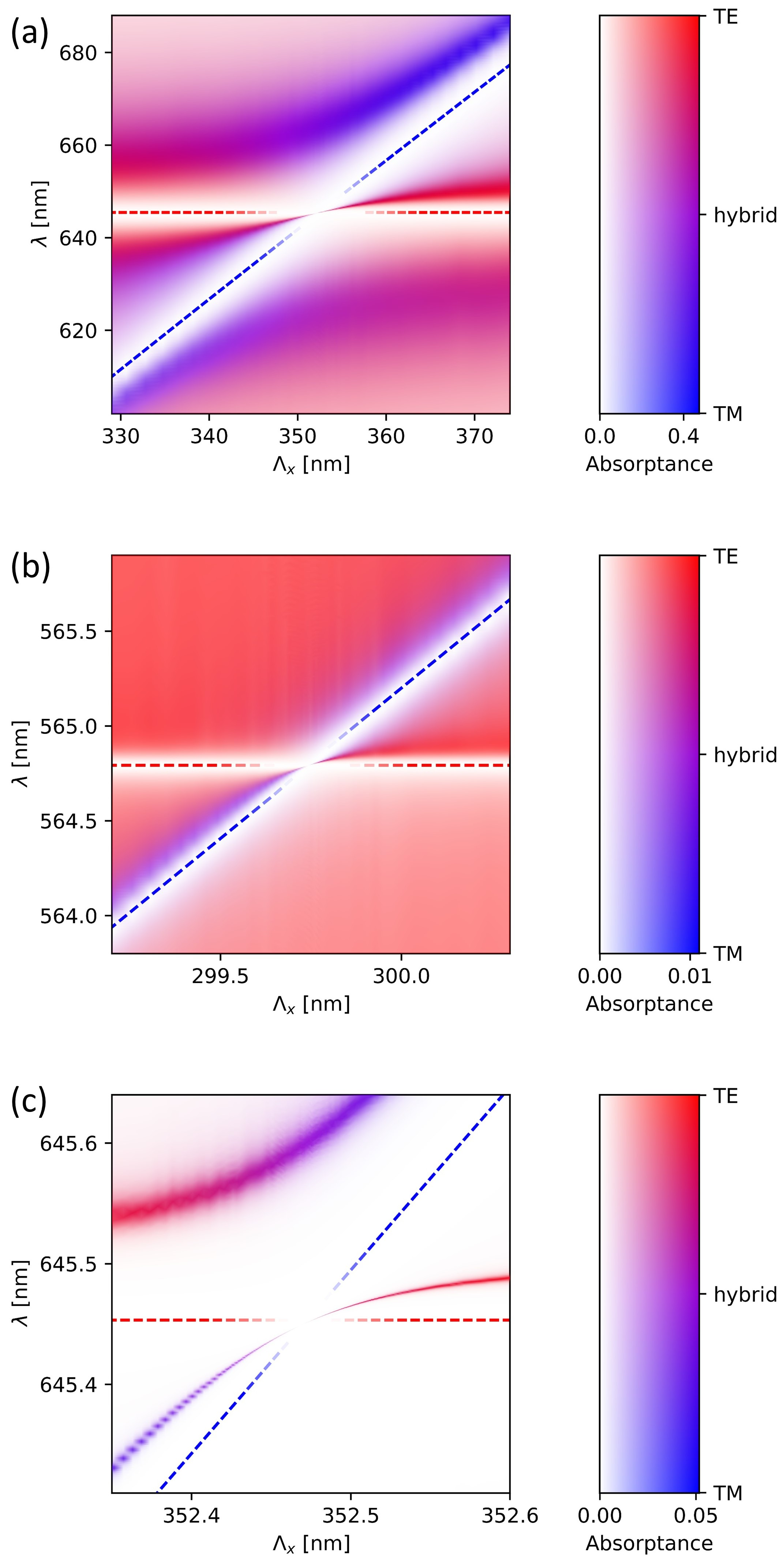}
\caption{\label{fig:3} Hybrid quasi-BICs revealed in metasurfaces with various geometrical parameters. The absorptance and the average polarization state of the excited electric field are plotted as functions of period $\Lambda_x$ (horizontal axis) and incident wavelength $\lambda$ (vertical axis). The absorptance is encoded in the opacity of the data points (displayed on a white background), whereas the polarization is encoded in the colors: red for the pure TE polarization, blue for the pure TM polarization, and violet for the hybrid polarization. The red and blue dashed lines illustrate the Bragg condition for the TE and TM guided modes, respectively. These lines are made transparent near the quasi-BICs not to obscure them. The graph in (a) corresponds to metasurfaces with nanocubes of size 75 nm, whereas in the remaining graphs, the size of nanocubes is set to 25 nm. In (b), $\Lambda_y = $ 292 nm, whereas in (a) and (c), $\Lambda_y = $ 340 nm. In (a) and (b), the quasi-BIC overlaps with the LSPR (a resonant scenario), whereas in (c), the quasi-BIC is away from the LSPR (an off-resonant scenario). The values of absorptance and polarization state were obtained using COMSOL Multiphysics. }
\end{figure}

\section{A plasmonic metasurface}

To demonstrate the capabilities of the proposed BIC, we consider a periodic metasurface composed of a 2D array of metal nanoparticles on the surface of a lossless slab waveguide [see Fig. \ref{fig:2}(a)]. We assume that the metasurface is sufficiently large, such that it can be described using periodic boundary conditions with the in-plane momentum $\mathbf{k}_{||}$. Hybrid optical metasurfaces of this type have been studied previously in the context of plasmonic-photonic resonances known as waveguide-plasmon polaritons (WPPs) \cite{christ03,rodriguez12}. In this work, we investigate the coupling between the TE and TM modes mediated by an array of lossy nanoparticles, giving rise to a hybrid BIC with suppressed absorption loss rate and a mixed TE-TM polarization state. Obviously, there can be found other coupling configurations for this demonstration, e.g., coupling between TE modes in a square lattice, or coupling between two counter-propagating modes in a simple one-dimensional grating. However, the purpose of the example we have chosen is to show that the proposed BIC can be formed independently of any specific symmetry requirement. The two modes coupled by metal nanoparticles in our example have different polarizations and different field profiles. Moreover, the structure under consideration lacks the up-down mirror symmetry that is required by the Friedrich-Wintgen BICs~\cite{lee20,glowadzka21}.

In the numerical calculations, the metal nanoparticles are gold nanocubes with the rib size ranging from 25 to 75 nm on the surface of a 300 nm thick waveguide with a Si$_3$N$_4$ core and a SiO$_2$ substrate. The optical constants of Si$_3$N$_4$, SiO$_2$, and Au are taken from Refs. \cite{luke15}, \cite{malitson65}, and \cite{johnson72}, respectively. In fact, the nanoparticle size, shape, and material composition are not critical for the emergence of the proposed BIC and can be chosen arbitrarily. However, we intentionally select nanoparticles that exhibit an electric-dipole-like localized surface plasmon resonance (LSPR) in the visible spectral range, which makes them strongly polarizable by the in-plane polarized evanescent fields of the guided modes. We also deliberately choose gold instead of silver as the nanoparticle material, as it shows significant optical absorption in the visible spectral range. This allows us to demonstrate the efficiency of our approach. In addition, the metasurface design is to some extent driven by the feasibility of its experimental realization and its possible future applications in spectroscopy.

Let us first consider a slab waveguide with a periodic perturbation of infinitesimal strength, i.e., a virtually periodic waveguide in the absence of the nanoparticles. Such a waveguide supports mutually orthogonal TE and TM guided modes described by the effective mode indices $n_{\sub{TE}}$ and $n_{\sub{TM}}$. Due to the periodicity of the system, each of the guided modes forms an onset of standing waves at the $\Gamma$ point ($|\mathbf{k}_{||}|=0$). These standing waves correspond to the band edges of the diffractive resonances with resonance frequencies $\omega$ governed by the 2$^{\sub{nd}}$ Bragg condition ($\omega=2\pi c/n\Lambda$ for the mode index $n$ and lattice period $\Lambda$). The resonance frequencies for the TE and TM modes can be matched by tuning the periods along the two lattice directions ($x$ and $y$), e.g., such that $n_{\sub{TE}}\Lambda_y = n_{\sub{TM}}\Lambda_x$. By looking at the electric field distributions in Figs. \ref{fig:2}(b) and \ref{fig:2}(c), one can clearly see that the TM and TE standing waves formed along the $x$ and $y$ axes, respectively, are both polarized parallel to the $x$-axis at the center of the unit cell on the waveguide surface. Hence, both these modes can resonantly couple to each other through surface-mounted nanoparticles.

\begin{figure*}[hbt!]
\centering\includegraphics[width=12.5cm]{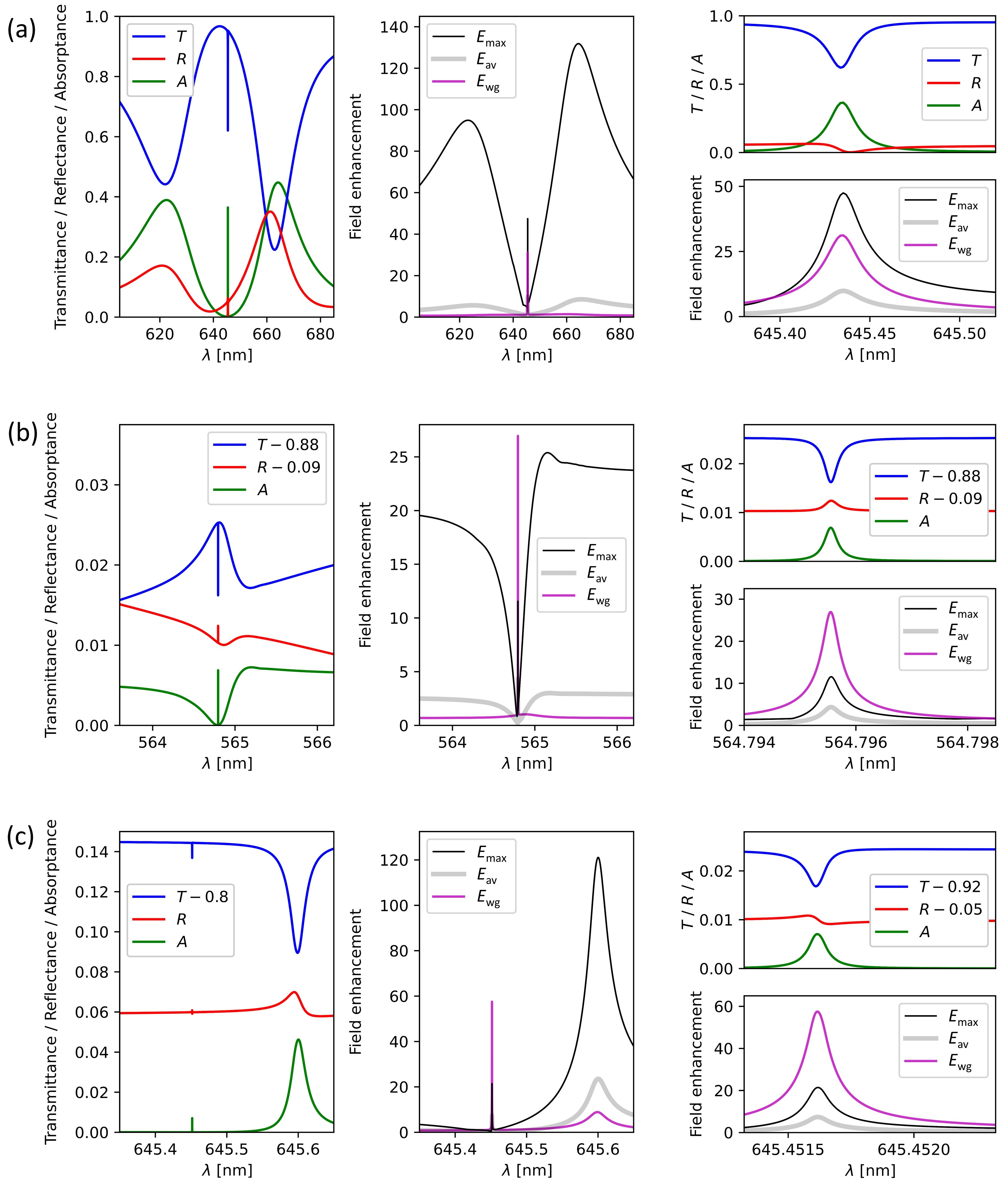}
\caption{\label{fig:4} Spectral dependence of the transmittance ($T$), reflectance ($R$), absorptance ($A$) (left column), and the values of the local field enhancement at the surface of the nanocubes (maximum $E_{\sub{max}}$ and average $E_{\sub{av}}$) and inside the waveguide ($E_{\sub{wg}}$) (middle column). The plots in (a), (b), and (c) reveal the quasi-BICs at $\Lambda_x = $ 352.5 nm, 299.75 nm, and 352.47 nm in Figs. \ref{fig:3}(a), \ref{fig:3}(b) and \ref{fig:3}(c), respectively. Graphs in the right column show the magnifications around the quasi-BICs of the graphs in the left and middle column. The values of transmittance, reflectance, absorptance and field enhancement were obtained using COMSOL Multiphysics. }
\end{figure*}

\begin{figure*}[hbt!]
\centering\includegraphics[width=15.5cm]{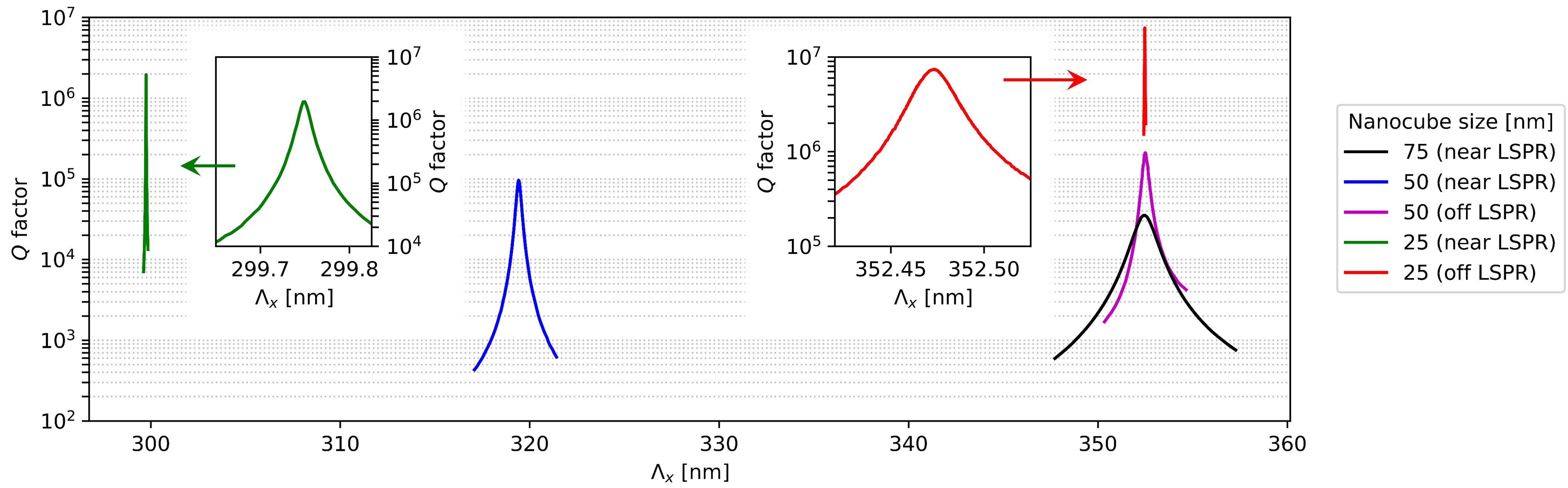}
\caption{\label{fig:5} $Q$ factors of the hybrid quasi-BICs as a function of $\Lambda_x$ in metasurfaces of various parameters (see legend). The black, green, and red curves correspond to the metasurfaces in Figs. \ref{fig:3}(a), \ref{fig:3}(b), and \ref{fig:3}(c), and Figs. \ref{fig:4}(a), \ref{fig:4}(b), and \ref{fig:4}(c), respectively. $\Lambda_y$ is set to 340 nm for the black, magenta, and red curves, 310 nm for the blue curve, and 292 nm for the green curve. The curves for 25 nm nanocubes (green and red lines) are magnified in the insets. All values were extracted from the spectral FWHM of the absorptance peaks calculated using COMSOL Multiphysics.}
\end{figure*}

The hybridization of the TE and TM standing waves mediated by the nanoparticles gives rise to three new resonances, including a hybrid dark resonance shown in Fig. \ref{fig:2}(d). Although in total, there are eight TE and TM standing waves at the $\Gamma$ point, the contribution of the other modes can be neglected. Hence, the description of the hybrid metasurface under consideration can be based on the Hamiltonian $\bm{H}$ of Eq. (\ref{eq:H}). The TE and TM standing waves can also individually hybridize with the nanoparticles when the system is tuned away from the TE-TM matching condition. In such a case, the hybridization gives rise to simple WPPs.

Figure \ref{fig:3} shows the optical properties of the hybrid metasurfaces at normal incidence as a function of period $\Lambda_x$ and wavelength $\lambda$. In particular, Fig. \ref{fig:3}(a) shows the anticrossing of the simple TE and TM WPPs (red and blue bands, respectively), resulting in hybrid TE-TM branches (violet) and a high-$Q$ BIC between them. In this case, the intersection of the Bragg conditions for the TE and TM modes (red and blue dashed lines, respectively) overlaps with the LSPR, which makes this scenario similar to that of Fig. \ref{fig:1}(b). Due to the relatively large size of the nanoparticles in this example (75 nm), the electric field of the BIC cannot completely avoid the nanoparticles. Despite this, we obtain a rather high $Q$ factor of 35000, as determined from the full width at half maximum (FWHM) of the absorptance peak in Fig. \ref{fig:4}(a). The peak originates from an efficient trapping and strong enhancement of the incident field by the metasurface at the BIC wavelength. The absorption loss rate of light from the system is still highly suppressed, providing a high $Q$ factor. For smaller nanoparticles [Figs. \ref{fig:3}(b) and \ref{fig:3}(c)], the $Q$ factor can be increased well above $10^6$ (up to $2.0 \times 10^6$ in the resonant case, and $7.4 \times 10^6$ in the off-resonant case), as determined from the peaks in Figs. \ref{fig:4}(b) and \ref{fig:4}(c). The values of $Q$ factors for various sets of parameters are presented in Fig. \ref{fig:5}. The $Q$ factor remains finite as long as the nanoparticles do not behave as point dipoles. Furthermore, the efficiency of loss cancellation in a real system depends on the validity of the approximations used to derive the Hamiltonian $\bm{H}$ in Eq. (\ref{eq:H}) from Eq. (\ref{eq:osc}). This validity can be affected, e.g.,  by the contribution of other modes (apart from the three modes considered here), or by the presence of unwanted coupling between the two lossless modes (other than the coupling mediated by the lossy mode). This means that the realizable BICs are \emph{quasi-BICs}, that are able to couple to the incident light, resulting in narrow spectral features, as can be seen in Fig. \ref{fig:4}.

The field enhancement spectra in Fig. \ref{fig:4} clearly show that, apart from ultrahigh $Q$ factors, the quasi-BICs can also produce a relatively high local field enhancement, both at the nanoparticle surface and in the waveguide. This is because the high-$Q$ resonance associated with the quasi-BIC allows the light to be efficiently trapped in the metasurface. As a result, its amplitude builds up via constructive interference in all locations, including the lossy nanoparticles that it tries to avoid. Consequently, the interaction of light with the nanoparticles may be strongly enhanced within a narrow resonance band~\cite{wang20,xiao21,saadabad21,zong23}. These properties make the presented quasi-BICs very promising for applications in optical sensing \cite{stewart08,conteduca22} and strong light-matter coupling \cite{torma14}. We emphasize that, in practice, ultrahigh $Q$ factors can be deteriorated by fabrication imperfections, finite illumination area (i.e., finite width of the angular spectrum of the incident beam) \cite{zundel22}, and finite lateral extent of the metasurface (side leakage) \cite{zou04,zundel18,jin19}. However, the proposed BICs are rather insensitive to the geometry of metal nanoparticles, which makes them robust and experimentally feasible compared to typical symmetry-enabled BICs \cite{sadrieva17}.

\section{Summary and conclusions}

To summarize, we have demonstrated the possibility of simultaneous suppression of the radiation and absorption losses in a plasmonic metasurface, creating BICs with $Q$ factors as high as $10^6$ - $10^7$, that can be further increased by optimizing the structure (e.g., replacing gold with silver). The presented mechanism of loss cancellation can be implemented in any system of two initially uncoupled and lossless oscillators interacting with a third oscillator that can have arbitrary losses. In this case, the condition for achieving an absorption-free BIC corresponds to the exact matching between the resonance frequencies of the two lossless oscillators. In the example system -- a slab waveguide and an array of plasmonic nanoparticles -- the TE and TM guided modes played the role of the two lossless oscillators, while the LSPR of the plasmonic nanoparticles served as the third lossy oscillator. The presented mechanism is not limited to the above specific example and can be realized in other physical systems within and beyond optics. 

In the context of optical BICs, simultaneous elimination of both the radiation and absorption losses could open up many possibilities. In plasmonics, for example, it could enable high-$Q$ resonances with strong local field enhancement, improving the performance of plasmonic systems \cite{liang20,bin21,bin22}. On the other hand, high-index dielectrics (e.g., Si, Ge, and GaAs), which are strongly absorbing in the visible spectral range, could be used to create high-$Q$ Mie-resonant photonic structures \cite{koshelev20}. This can be very useful in all applications relying on resonant enhancement of light-matter interaction, including ultrasensitive spectroscopic devices, nonlinear optical modulators and frequency converters, and light emitters with tailored characteristics \cite{vaskin19,hwang22,kolkowski23}. 

\section*{Acknowledgments}

The authors acknowledge the support of the Academy of Finland (Grants No. 347449 and 353758) and the Flagship of Photonics Research and Innovation (PREIN) funded by the Academy of Finland (Grant No. 320167). For computational resources, the authors acknowledge the Aalto University School of Science “Science-IT” project and CSC – IT Center for Science, Finland.

\section*{Disclosures}

The authors declare no conflicts of interest.

\section*{Data Availability}

Data underlying the results presented in this paper and their replication instructions are openly available at \href{https://doi.org/10.23729/cfe98559-2ca5-43c6-b8f3-b542f9ff94bb}{https://doi.org/10.23729/cfe98559-2ca5-43c6-b8f3-b542f9ff94bb}, Ref. \cite{10.23729/cfe98559-2ca5-43c6-b8f3-b542f9ff94bb}.

\invisible{.}

\bibliography{lib}

\end{document}